\documentstyle[preprint,aps,epsfig,amssymb]{revtex}
\def\br{\begin{eqnarray}}
\def\er{\end{eqnarray}}
\def\be{\begin{equation}}
\def\ee{\end{equation}}

\def\L{\Lambda}

\def\({\left(}
\def\){\right)}
\def\<{\left\langle}
\def\>{\right\rangle}

\begin{document}
%
%
\title{Strength of the Trilinear Higgs Boson Coupling in Technicolor Models}
\author{
A. Doff and A. A. Natale\\
}
\address{
Instituto de F\'{\i}sica Te\'orica, UNESP,
Rua Pamplona 145,
01405-900, S\~ao Paulo, SP,
Brazil}
\date{\today}
\maketitle
\begin{abstract}
We discuss the strength of the trilinear Higgs boson coupling in technicolor (or composite)
models in a model independent way. The coupling is determined as a function of a very general ansatz for the technicolor self-energy, and turns out to be equal or smaller than the one of the standard model Higgs boson depending on the dynamics of the theory. With this trilinear coupling we estimate the cross section for Higgs boson pair production at the LHC. This measurement is quite improbable in the case of a heavy standard model Higgs boson, but it will be even worse when this boson is dynamically generated.
\end{abstract}



\section{Introduction}

In the standard model of elementary particles the fermion and gauge boson
masses are generated due to the interaction of these particles with elementary
Higgs scalar bosons. Despite its success there are some points in the model
as, for instance, the enormous range of masses between the lightest and heaviest
fermions and other peculiarities that could be better explained at a deeper level. The nature of the Higgs boson is one of the most important problems in particle physics, and there are many questions that may be answered in the near future by the LHC experiments, such as: Is the Higgs boson, if it exists at all, elementary or composite? What are
the symmetries behind the Higgs mechanism?

\par Among the priorities of the Higgs boson search at the LHC experiments, is the
measurement of its mass, width, spin and CP eigenvalues. The measurement of the Higgs
boson couplings and, particularly, its self-couplings will also be quite important,
once they may unravel all the subtleties of the mechanism of electroweak symmetry 
breaking \cite{gunion}. It may be possible
to measure the Higgs boson self-coupling at the LHC in the case of a light Higgs boson\cite{baur}
, but this will barely be possible if the boson mass is larger than $200$ GeV.

There are many variants for the Higgs mechanism. Our interest in this work will be
focused in the models of  electroweak symmetry breaking via  strongly interacting theories of technicolor type \cite{lane}. In these theories the Higgs boson is a composite
of the so called technifermions, and at some extent any model where the
Higgs boson is not an elementary field follows more or less the same ideas of the 
technicolor models. In extensions of the standard model the scalar
self-couplings can be enhanced, like in the supersymmetric version\cite{zerwas}. If the same 
happens in models of dynamical symmetry breaking, as far as we know, has not been investigated
up to now.

The beautiful characteristics of technicolor (TC) as well as its problems were clearly listed recently by Lane\cite{lane,hs}. Most of the technicolor
problems may be related to the dynamics of the theory as described in
Ref.\cite{lane}. Although technicolor is a non-Abelian gauge theory it is
not necessarily similar to QCD, and if we cannot even say that QCD is fully understood
up to now, it is perfectly reasonable to realize the enormous work that is needed
to abstract from the fermionic spectrum the underlying technicolor dynamics.

The many attempts to build a realistic model of dynamically generated
fermion masses are reviewed in Ref.\cite{lane,hs}. Most of the work in this area
try to find the TC dynamics dealing with the particle content of the
theory in order to obtain a technifermion self-energy that does not
lead to phenomenological problems as in the scheme known as walking
technicolor\cite{walk}. The idea of this scheme is quite simple. First,
remember that the expression for the TC self-energy is
proportional to   $ \Sigma (p^2)_{{}_{TC}} \propto { (\langle \bar{\psi} \psi\rangle_{{}_{TC}}}/{p^2})
(p^2/\L^2_{{}_{TC}})^{\gamma^*}$, where $\langle \bar{\psi} \psi\rangle_{{}_{TC}}$ is
the TC condensate and $\gamma^*$ its anomalous dimension. Secondly, depending
on the behavior of the anomalous dimension we obtain different
behaviors for $ \Sigma (p^2)_{{}_{TC}}$. A large anomalous dimension may solve the
problems in TC models. In principle we could deal with
many different models, varying fermion representations and particle content,
finding different expressions for  $ \Sigma (p^2)_{{}_{TC}}$ and testing them
phenomenologically, i.e. obtaining the fermion mass spectra without
any conflict with experiment. Usually the walking behavior is
obtained only with a large number of technifermions, although there are recent proposals where 
the walking behavior is obtained for a very small number of fields with the introduction of technifermions in higher dimensional representations of the technicolor gauge group\cite{sannino}.
 
As the dynamics in models of dynamical
symmetry breaking can be so different from QCD, it is interesting to investigate the behavior
of the dynamical Higgs boson self-coupling. It could be possible that in these models
the self-coupling is enhanced and easy to measure than the standard one, even considering 
that the dynamical Higgs boson will be heavier than $200$ GeV.

In this work we will consider a very general ansatz for the technifermion self-energy that was introduced in the Ref.\cite{aa1}. This ansatz interpolates between all  known forms of technifermionic self-energy. As we vary some parameters in our ansatz for the technifermionic self-energy we go from the standard operator
product expansion (OPE) behavior of the self-energy to the one predicted by the extreme limit of a walking
technicolor dynamics, i. e. $\gamma^* \rightarrow 1 $ \cite{walk,soni,soni2}.We will discuss the general properties of the trilinear Higgs coupling based on this ansatz. In principle the trilinear Higgs self-coupling can be measured directly in the Higgs boson pair production at the LHC via the gluon-gluon fusion mechanism. 
In this way we can predict the general behavior of this cross section in technicolor models. As, in non-Abelian
gauge theories, the fermion self-energy is related to the Bethe-Salpeter wave-function of the composite scalar
boson \cite{delbourgo}, we believe that our result could also be extended to any composite Higgs model.   
 
\par This paper is organized as follows: In Sec. II we compute  the trilinear self-coupling of a composite Higgs boson assuming the anzatz for the fermionic self-energy shown in Ref.\cite{aa1}. In the Sec. III  we review the self-couplings of the standard model fundamental Higgs field and compare them with the results shown in the previous section. In the Sec. IV  we compute the cross section for the $gg \rightarrow HH$ reaction in the case of the standard model and for a composite Higgs boson. Finally in the Sec.V  we draw our conclusions.

\section{The Trilinear Self-Coupling for a Composite Higgs Boson }
 
\par Using Ward identities we can show the couplings of the scalar boson to fermions to be\cite{soni}
\be 
G^{\sl a} (p+q,p) = -\imath \frac{g_{W}}{2M_{W}}
\left[\tau^{\sl a}\Sigma(p)P_R - \Sigma(p+q)\tau^{\sl a} P_L \right]
\label{fsc}
\ee
where $P_{R,L} = \frac{1}{2} (1 \pm \gamma_5 )$, $\tau^{\sl a}$ is a $SU(2)$
matrix, and $\Sigma$ is a matrix of fermionic self-energies in weak-isodoublet
space. As in Ref.\cite{soni} we assume that there is a scalar composite Higgs
boson that couples to the fermionic self-energy which is saturated by the
top quark\cite{aa2}. Specifically, we assume that the scalar-to-fermion coupling matrix at large momenta is given by $G(p,p)$, where we do not attempt to distinguish between the two fermion momenta $p$ and $p+q$, since, in all situations with which we will be concerned, $\Sigma(p+q)\approx \Sigma(p)$. Therefore the coupling between a composite Higgs boson with fermions at large momenta is given by  
\be 
\lambda_{{}_{Hff}}(p)\equiv G(p,p) \sim -\frac{g_{W}}{2M_{W}}\Sigma(p^2)
\label{fh}
\ee 
where $\Sigma(p^2)$ is the fermionic self-energy. The trilinear Higgs boson coupling
in technicolor models will be dominated by loops of heavy fermions that couple to
the scalar Higgs particle as predicted by Eq.(\ref{fh}) \cite{soni}. Our purpose in this  section is to obtain  an expression for the trilinear Higgs boson coupling using the  ansatz  
\be 
\Sigma_{A}(p^2) \sim \Lambda_{{}_{TC}}\left( \frac{ \Lambda^2_{{}_{TC}}}{p^2}\right)^{\alpha}\left[1 + a\ln\left(p^2/ \Lambda^2_{{}_{TC}} \right) \right]^{-\beta}  \,\,\, ,
\label{sig}
\ee	
which was proposed in the Ref.\cite{aa1}. This choice interpolates between  the standard OPE result for the technifermion self-energy, which is  obtained when $\alpha \rightarrow 1$, and the extreme walking technicolor solution obtained when $\alpha \rightarrow 0$ \cite{walk}, i.e. this is the case where the symmetry breaking is dominated by higher order interactions that are relevant at or above the TC scale, leading naturally to a very hard dynamics \cite{soni,soni2}.
As we have pointed out in Ref. \cite{aa2} only such kind of solution  is naturally  capable of generating a large mass to the third fermionic generation, which has a mass limit almost saturated by the top quark mass.  Moreover, as also claimed in the second paper of Ref.\cite{aa2}, there are other possible reasons to have $\alpha \sim 0$, as the existence of an infrared fixed point and a gluon (or technigluon) mass scale \cite{alkofer}, which, actually, are related possibilities \cite{coup}.

\par The Yukawa top quark coupling to the Higgs boson is large, no
matter we are considering the composite or the fundamental standard model Higgs boson, and
is the one that dominates the process that we will be considering in the next sections. The
main difference is that in the composite case the trilinear coupling is a function of this
Yukawa coupling as remarked in Ref.\cite{soni,soni2}. As considered in Ref.\cite{soni} and many
others dealing with dynamical symmetry breaking models it is usually assumed that such calculations
are not spoiled by higher order corrections. It is interesting that many technicolor models
make use of the existence of a non-trivial fixed point (or a quasi-conformal theory) to
cure their phenomenological problems\cite{walk}, and exactly for this
possibility Brodsky has been claiming that it will be possible to built an 
skeleton expansion that could allow to capture the non-perturbative
effects in a reliable way\cite{brodsky}.  

\par In the Eq.(\ref{sig}) the scale, $\Lambda_{{}_{TC}}$ is related to the technicolor condensate by 
$\langle \bar{\psi} \psi\rangle_{{}_{TC}} \approx \Lambda^{3}_{{}_{TC}}$. We defined $\beta \equiv \gamma_{{}_{TC}}\cos(\alpha\pi)$, $a \equiv bg^2_{{}_{TC}}$ with $\gamma_{{}_{TC}} = 3c/16\pi^2b$,  and  $c$ is the quadratic Casimir operator given by 
$$ 
c = \frac{1}{2}\left[C_{2}(R_{1}) +  C_{2}(R_{1}) - C_{2}(R_{3})\right]
$$ 
where $C_{2}(R_{i})$,  are the Casimir operators for technifermions in the representations  $R_{1}$ and 
$R_{2}$ that condensate in the representation $R_{3}$, $b$ is the coefficient of the $g^3$ term in the technicolor $\beta (g)$ function. 
\par We can determine one expression for the trilinear coupling for any theory where the Higgs boson is  composite  by considering the diagram shown in Fig.(1).
\par
\begin{figure}[h]
\begin{center}
\epsfig{file=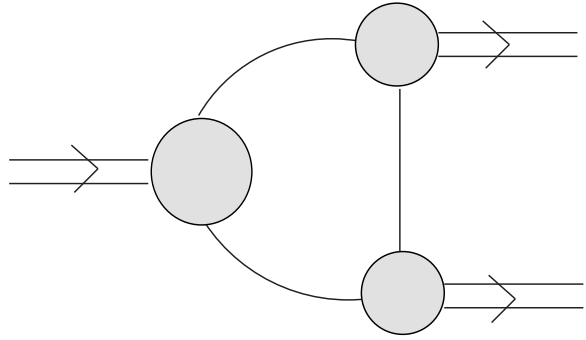,width=0.5\textwidth}
\caption{The gray blobs in this figure represent the coupling of composite Higgs bosons to fermions. The double lines represent the composite Higgs bosons. The full diagram is the main contribution to the trilinear Higgs boson self-coupling.}
\label{fig1}
\end{center}
\end{figure}

 The contribution of Fig.(\ref{fig1}) is certainly the dominant one \cite{soni}. Assuming the coupling of the scalar boson to the fermions given by the Eq.(1), and with the fermion propagator written as 
\be 
S_{F}(p) = \frac{(\not\!p + \Sigma(p^2))}{(p^2 - \Sigma^2(p^2))}
\ee  
we find that 
\be 
\lambda^{{T}}_{{}_{HHH}} = \frac{3g^3_{W}}{64\pi^2}\left(\frac{3n_{F}}{M^3_{W}}\right)\int^{\infty}_{0}\!\!\!\!\frac{\Sigma^4(p^2)p^4dp^2}{(p^2 + \Sigma^2(p^2))^3}.
\label{tri} 
\ee 
\noindent where  $n_{F}$ is the number of technifermions included in the model. Considering the ansatz given by the Eq.(\ref{sig}),  and introducing it into Eq.(\ref{tri}), we obtain 
\be 
\lambda^{T}_{{}_{HHH}} \approx  \frac{3g^3_{W}}{64\pi^2}\left(\frac{3n_{F}}{M^3_{W}}\right)\Lambda^4_{{}_{TC}}(\Lambda_{{}_{TC}})^{4\alpha}I(p^2)
\label{eq5}
\ee 
with
$$ 
I(p^2) = \frac{1}{\Gamma(4\beta)}\int^{\infty}_{0}\!\!\!dzz^{4\beta - 1}e^{-z}(\Lambda_{{}_{TC}})^{az}\int^{\infty}_{0}\!\!\!\frac{dp^2(p^2)^{2-4\alpha-az}}{(p^2 + \Lambda^2_{{}_{TC}})^3}
$$ 
To compute  this last expression we have used the following Mellin transform 
\be 
[1 + A\ln{B}]^{-\eta}= \frac{1}{\Gamma(\eta)}\int^{\infty}_{0}dzz^{\eta -1}e^{-z}(B)^{-Az}.
\ee  

\par After performing the $p^2$ integration in Eq.(\ref{eq5}), we  can write this equation as 
\be 
\lambda^{T}_{{}_{HHH}} \approx  \frac{3g^3_{W}}{64\pi^2}\left(\frac{3n_{F}}{M^3_{W}}\right)\frac{\Lambda^4_{{}_{TC}}}{\Gamma(4\beta)}\int^{\infty}_{0}\frac{dzz^{4\beta - 1}e^{-z}}{4\alpha + az}.
\label{eq7}
\ee 
\par We will present our analysis of $\lambda^{T}_{{}_{HHH}}$ for two different regions of the parameter $\alpha$. We will start with the case  $\alpha \approx 0$. Therefore we can make the following expansion in Eq.(\ref{eq7})
\be
\frac{1}{4\alpha + az} \approx \frac{1}{az}\left[ 1 - \frac{4\alpha}{az} + O(\alpha^2) ... \right].  
\ee  
Than Eq.(\ref{eq7}) can be cast in the form 
\br
\lambda^{T0}_{{}_{HHH}}  \approx  \frac{3g^3_{W}}{64\pi^2}\left(\frac{3n_{F}}{M^3_{W}}\right)&&\frac{\Lambda^4_{{}_{TC}}}{a\Gamma(4\beta)}\left[ \int^{\infty}_{0}dzz^{4\beta - 2}e^{-z} + \right. \nonumber \\  &&\!\!\!\!\!- \left. \frac{4\alpha}{a}\!\! \int^{\infty}_{0}\!\!\!\!\! dzz^{4\beta - 3}e^{-z} + O(\alpha^2)...\right] \nonumber   
\er  
Retaining only the first two terms in the $\alpha$ expansion and performing the $z$ integration, we finally can write 
\be 
\lambda^{T0}_{{}_{HHH}} \approx  \frac{3g^3_{W}}{64\pi^2}\left(\frac{3n_{F}}{M^3_{W}}\right)\frac{\Lambda^4_{{}_{TC}}}{a(4\beta -1)}\left[1 - \frac{4\alpha}{a}\frac{1}{(4\beta - 2)}\right].
\label{eq9}
\ee 
\par When $\alpha \approx 1$, we can consider a similar expansion, and following the same steps we obtain 
\be 
\lambda^{T1}_{{}_{HHH}} \approx  \frac{3g^3_{W}}{64\pi^2}\left(\frac{3n_{F}}{M^3_{W}}\right)\frac{\Lambda^4_{{}_{TC}}}{4}\left[1 - \frac{4}{a}(\alpha - 1)\right]
\label{eq10}
\ee  
The above expressions for the trilinear Higgs coupling are quite dependent on the scale $\Lambda_{{}_{TC}}$. This is not the best formula to compute this coupling, since $\Lambda_{{}_{TC}}$, which in principle is related to the value of the dynamical technifermion mass at the origin, is not directly fixed by the symmetry breaking of the standard
model. A more appropriate quantity that can be used to describe this coupling is the technipion decay constant, which is fixed by the $W$ and $Z$ gauge boson masses.  

Considering our comments in the previous paragraph we will express the trilinear Higgs coupling as a function
of the technipion decay constant ($F_{{}_{\Pi}}$) instead of the scale $\Lambda_{{}_{TC}}$. 
$F_{{}_{\Pi}}$ can be computed through the known Pagels and Stokar relation\cite{pagels} 
\be
F^2_{\Pi} = \frac{N_{{}_{TC}}}{4\pi^2}\!\!\!\int^{\infty}_{0}\!\!\!\!\!\!\frac{dp^2p^2}{(p^2 + \Sigma^2(p^2))^2}\!\!\left[\Sigma^2(p^2) - \frac{p^2}{2}\frac{d\Sigma(p^2)}{dp^2}\Sigma(p^2)\right]
\ee 
\noindent where $N_{{}_{TC}}$ is the technicolor number. 
\par  We compute the technipion decay constant using the ansatz Eq.(\ref{sig}). After some calculation we  obtain the following expression for $F_{{}_{\Pi}}$
\be 
F^2_{{}_{\Pi}} = \frac{N_{{}_{TC}}}{4\pi}\Lambda^2_{{}_{TC}}f(k)
\label{eq12}
\ee 
where   
\be 
f(k) = \frac{(1 + k/2)}{(1 + 2k)^2}csc\left[ \pi/(1 + 2k)\right]
\ee 
with 
$$ 
k = \alpha + 3\cos(\alpha\pi)/4\pi
$$
To obtain this expression we have assumed the scaling law $c\alpha_{{}_{TC}} \sim 1$\cite{rabi}. To be consistent with  Eqs.(\ref{eq9}) and (\ref{eq10}), we  also need to expand Eq.(\ref{eq12}) for $\alpha \approx 0$ and $\alpha \approx 1$. In this case, we obtain 
\be 
F^2_{{}_{\Pi}} = \frac{N_{{}_{TC}}}{8\pi}\Lambda^2_{{}_{TC}}\left[1 - S(\alpha)\right] 
\ee
with 
$$ 
S(\alpha) = \left\{ \begin{array}{c} 5\alpha \,\,\,\,{\rm for}\,\,\,\, \alpha\approx 0 \\ \\ \alpha/2 \,\,\,\,{\rm for}\,\,\,\, \alpha\approx 1  \end{array}\right.
$$
\par Finally, assuming this last equation, we can write  the Eqs.(\ref{eq9}) and  (\ref{eq10}) in the form 
\be 
\lambda^{T\alpha}_{{}_{HHH}}= 3n_{F}\frac{F_{{}_{\Pi}}}{N^2_{{}_{TC}}}f(\alpha)
\label{eq15}
\ee  
where for convenience we  defined
$$ 
f(\alpha) = \left\{ \begin{array}{c} \frac{3}{a(4\beta - 1)}\frac{[1 - 4\alpha/a(4\beta - 2) ]}{(1 - 5\alpha)^2}\,\,\,\,{\rm when}\,\,\,\, \alpha\approx 0 \\ \\ \frac{3}{4}\frac{[1 - 4(\alpha - 1)/a ]}{(1 - \alpha/2)^2} \,\,\,\,{\rm when}\,\,\,\, \alpha\approx 1  \end{array}\right.
$$ 
and will assume $F_{{}_{\Pi}} = 125 GeV.$\footnote{In TC models containing $N_{{}_{D}}$ doublets of technifermions $F_{{}_{\Pi}} = 250GeV/\sqrt{N_{{}_{D}}}$,  and in this work we will be assuming  $N_{{}_{D}}=4$.}
\par Our ansatz for the fermionic self-energy is a very general one. No matter which is
the theory (technicolor or any of its variations) the self-energy will be limited
to the expressions obtained from Eq.(\ref{sig}) for $\alpha$ in the range $[0,1]$,
even the scenario proposed in Ref.\cite{sannino} will be described by such 
expression.
\par In the next section we will compare these expressions for the trilinear composite Higgs boson self-coupling with the one of the standard model fundamental Higgs boson.

\section{Trilinear Coupling: Fundamental $\times$ Composite Higgs boson} 

In this section  we review the expression for the trilinear coupling in the case of the standard model fundamental Higgs boson, and compare it to the ones found in the previous section. We start writing the expression of the Higgs boson potential in the  Standard Model 
\be 
V(\varphi) = -\mu^2\varphi^{\dagger}\varphi + \lambda(\varphi^{\dagger}\varphi)^2.
\ee 
\noindent The self-couplings are uniquely determined in the Standard Model by the mass of the Higgs boson, which is related to the  quadrilinear coupling $\lambda$  by the following expression 
$$ 
M^2_{H} =  2\lambda v^2.
$$ 
\par After introducing the physical Higgs field H in the neutral component of the doublet 
$\langle\varphi\rangle = (v +H)/\sqrt{2}$ we can write the potential as
\be 
V(H) = \frac{M^2_{H}}{2}H^2 +  \frac{M^2_{H}}{2v}H^3 + \frac{M^2_{H}}{8v^2}H^4.
\ee  
\noindent The multiple Higgs couplings can be derived from the potential $V(H)$, and the trilinear and quadrilinear couplings of the Higgs field H are given by  
\br 
&&\lambda_{{}_{3H}}= 3\frac{M^2_{{}_{H}}}{M^2_{{}_{Z}}}\lambda_{0} \nonumber \\ 
&&\lambda_{{}_{4H}}= 3\frac{M^2_{{}_{H}}}{M^4_{{}_{Z}}}\lambda^2_{0} \,\, .
\label{e18}
\er  
To obtain these expressions we assumed the normalization employed in Ref.\cite{zerwas}, where $\lambda_{0} = M^2_{{}_{Z}}/v$. 

In the case of a composite Higgs boson it is possible to show that its mass can be expected to be of the following
order \cite{delbourgo}:
$$
M_{{}_{H}} \sim 2\Lambda_{{}_{TC}}  \,\,\, .
$$ 
This result is independent of the dynamics and is originated from the similarity between the Schwinger-Dyson equation for the technifermion self-energy and the Bethe-Salpeter equation for the scalar channel \cite{delbourgo}. Of course, as discussed in the previous section, we will write $M_{{}_{H}}$ as a function
of $F^2_{{}_{\Pi}}$ instead of $\Lambda_{{}_{TC}}$.
\par To compare the results of the previous section with the couplings shown above we  can write the couplings for the composite Higgs boson as a function of its mass. Assuming the mass relation given above, considering the Eq.(\ref{eq15}) and rewriting it in terms of the parameter $\lambda_{0}$, we obtain in the case of  $\alpha = 0$
\be 
\lambda^{T0}_{{}_{3H}} = \left(\frac{1}{14}\right) \frac{n_{{}_{F}}M_{{}_{H}}}{N_{{}_{TC}}\sqrt{2\pi N_{{}_{TC}} }}\frac{\hat{\lambda}_{0}}{a(4\beta - 1)} \,\, ,
\label{e19}
\ee 
\noindent  and in the case when $\alpha = 1$ we obtain
\be 
\!\!\!\!\!\!\!\!\!\!\!\!\!\!\!\!\!\!\!\!\!\!\lambda^{T1}_{{}_{3H}} = \left(\frac{1}{28}\right) \frac{n_{{}_{F}}M_{{}_{H}}}{N_{{}_{TC}}\sqrt{\pi N_{{}_{TC}}}}\hat{\lambda}_{0}.
\label{e20}
\ee 
\noindent where $\hat{\lambda}_{0} \equiv \lambda_{0}/( 1 \,\, GeV)$.
\par In the Fig.(2) the behavior of the trilinear Higgs couplings is plotted as a function of the Higgs boson mass. The solid line represents the contribution of the fundamental Higgs boson, i. e. the Standard Model Higgs boson.
\par The dynamics of the extreme limit of a walking technicolor theory will be responsible for a fermionic self-energy that is given by the limit $\alpha \rightarrow 0$ in Eq.(\ref{sig}). To compare the trilinear Higgs coupling for fundamental and composite scalar bosons we will consider technicolor models with technifermions in the fundamental representation and will choose appropriately the number
($n_{F}$) of technifermions in order to obtain the desired walking behavior. For example, 
if the technicolor group is $SU(2)_{{}_{TC}}$, the walking limit is going to be obtained
with $n_{F} = 8$. The 8 technifermions can be reconized as a colored weak doublet $Q = (U^{a}, D^{a})$\footnote{In this expression $a=1..3$ is a color index.}, and a color-singlet weak doublet $L=(E, N)$. If the technicolor theory is described by 
the $SU(4)_{{}_{TC}}$ non-Abelian group, the extreme walking behavior is obtained when $n_{F}\sim 14$, which can be built with the addition of two colored  weak singlets $(R^a,S^a)$. Of course we are not discussing about phenomenologically viable models, but
the cases that we are presenting are plausible examples to make the comparison between
the ``composite" and the elementary coupling. 
\par In the Fig.(2) the continuous curve shows the behavior of the trilinear Higgs boson
self-coupling given by Eq.(\ref{e18}). In the same figure we indicate by 
$(\square,  \blacksquare)$ the  values of the trilinear composite Higgs  couplings
obtained respectively with the help of Eqs.(\ref{e19}) and (\ref{e20}) $(\alpha \rightarrow 0, \alpha \rightarrow 1)$ in the case of the $SU(2)_{{}_{TC}}$ technicolor group. 
\noindent We also indicate by  $(\vartriangle, \blacktriangle)$ in Fig.(2) the values of the trilinear coupling obtained for the $SU(4)_{{}_{TC}}$ when the parameter $\alpha$ has respectively the following behavior $(\alpha \rightarrow 0, \alpha \rightarrow 1)$.

It is possible to verify in Fig.(2) that the trilinear Higgs coupling generated by the 
dynamics in the limit $\alpha \rightarrow 0$, which corresponds to the extreme walking
technicolor limit, are quite close to the values obtained in the case of the fundamental
standard model Higgs boson. However, in the limit $\alpha \rightarrow 1$ the behavior
predicted for the trilinear Higgs coupling is very different; it decreases the more the
technicolor dynamics approaches the standard result predicted by simple OPE analysis. 
The arrow in Fig.(2) shows roughly the expected change in the trilinear coupling as we
go from $\alpha \rightarrow  0$ to $\alpha \rightarrow 1$. As will be shown in the next
section this behavior will imply in smaller cross section for Higgs boson pair production. 

\section{Cross Section for Higgs Boson pair production via gluon-gluon Fusion}

\par The large number of gluons in high-energy proton beams  imply that the gluon-gluon fusion mechanism is the dominant process for Higgs boson pair prodution at the LHC\cite{zerwas}. 
The gluon-gluon mechanism involves the triangular and box loops of the heavy quarks, as shown in Fig.(3), and is probably the best way to measure the trilinear Higgs coupling \cite{eboli}. 
\par In the Standard Model, only the top quark, and to a lesser extent the bottom quark, will contribute to the amplitudes of the diagrams in Fig.(3). In this section we will compute the twin Higgs boson production with the trilinear couplings that we obtained in the previous sections, in order to verify what this process can tell to us about the structure of the technicolor theory. 

\par In terms of the trilinear Higgs coupling, $\lambda_{{}_{3H}}= 3M^2_{{}_{H}}/M^2_{{}_{Z}}$, the elementary $g+g \, \rightarrow \, H+H$ cross section  at leading order can be written as\cite{zerwas}
\be 
\hat{\sigma} = \int^{\hat{t}^+}_{\hat{t}^{-}}\frac{G^2_{F}\alpha^2_{s}}{256(2\pi)^3}\left[|C_{T}F_{T} + C_{B}F_{B}|^2 + |C_{B}G_{B}|^2\right]d\hat{t}
\ee 
where the couplings $C_{T}$ and $C_{B}$ are defined as 
$$ 
C_{T} = \lambda_{{}_{HHH}}\frac{M^2_{Z}}{\hat{s} - M^2_{H}} \,\,\, ,\,\,\, C_{B} = 1, 
$$ 
with $\hat{s} = \tau s$ , $\hat{t}^{\pm} = -1/2[\hat{s} - 2M^2_{{}_{H}}\mp\hat{s}\beta_{H} ]$  and  
\be 
\beta_{H} = \sqrt{1 - 4M^2_{{}_{H}}/\hat{s}}.
\ee
The form factors $F_{T}$, $F_{B}$ and $G_{B}$ come from the triangle and box loop evaluation.
\par It is possible to derive simple expressions for the cross section in the limit where the Higgs boson mass is 
much lighter or much heavier than the internal quark ($q$) mass that is running in the loops. According to the discussion of the previous section it is possible to assume the limit where $M^2_{{}_{H}}\gg m^2_{q}$, because for $F_{{}_{\Pi}} = 125 GeV$ we can expect that $M_{{}_{H}} \sim O(1)TeV$\cite{ref19}. In this case the form factors $F_{{}_{T}}, F_{{}_{B}}$ and $ G_{{}_{B}}$ take a very simple form according to the last article quoted in the Ref.\cite{zerwas} 
\be
F_{{}_{T}} \sim -\frac{m^2_{q}}{\hat{s}}\left[\log\frac{m^2_{q}}{\hat{s}} + i\pi\right]\,\,\,,\,\,\,F_{{}_{B}} \sim G_{{}_{B}} \simeq 0,  
\ee
\noindent what allow us to write the partonic cross section as  
\be 
\hat{\sigma} = \frac{\alpha^2_{{}_{W}}\alpha^2_{s}}{4096\pi}\frac{M^4_{Z}}{M^4_{W}}\frac{\hat{s}\beta_{{}_{H}}}{(\hat{s} - M^2_{{}_{H}})^2}\lambda^2_{{}_{HHH}}|F_{{}_{T}}|^2.
\ee 
\noindent The total cross section for Higgs boson pairs production  through gluon-gluon fusion in pp collisions can be determined by integrating over the $gg$ luminosity\cite{zerwas}
\be 
\sigma_{T}(pp \rightarrow HH) = \int^{1}_{4M^2_{{}_{H}}/s}d\tau\frac{d{\cal{L}}^{gg}}{d\tau}\hat{\sigma}(\hat{s}=\tau s)
\ee 
where 
\be 
\frac{d{\cal{L}}^{gg}}{d\tau} = \int^{1}_{\tau}\frac{dx}{x}g(x,\mu^2)g(\tau/x,\mu^2)
\ee 
\noindent 
and we use the gluons distribution function ($g(x,\mu^2)$) obtained in Ref.\cite{martin}, taken at a typical scale $\mu \sim M_{{}_{Z}}$.
The cross section contains terms proportional to $m_q^2/s$ (where 
$m_q$ is the mass of the quark running in the loop) which were expanded and only the
leading term was considered. This is reasonable because the heaviest quark mass is the one
of the top quark and we are considering heavier scalar masses.  These equations have been numerically integrated and in the sequence we plot the  cross-section  as a function of the Higgs boson mass.
\par In Fig.(4) the curve shows the standard model prediction for Higgs boson pair production. We are also pointing out four values of the cross-section for the cases
of composite Higgs boson that were discussed in the previous section. The points
indicated by  ($\square, \blacksquare$) are the predictions for the $SU(2)_{{}_{TC}}$
technicolor group respectively in the limits $\alpha = 0$ and $\alpha = 1$.
The composite Higgs boson mass is estimated as being of order $M^{dyn}_{{}_{H}} \sim O(0.9 - 1.2)TeV$ as we vary the value of $\alpha$ in the range $0-1$. In the same figure we are also considering the case where $SU(4)_{{}_{TC}}$ is the TC group, again its
predictions are indicated by ($\vartriangle, \blacktriangle$). The Higgs mass obtained  in this case is $M^{dyn}_{{}_{H}} \sim O(600 - 900)GeV$. To compute these cross sections  we used the top quark mass equal to $m_{t} = 178GeV$.  
\par The cross section falls with increasing $M_{{}_{H}}$ due to both the diminishing luminosity and the decrease of the phase space. According to what we commented  in the end of Sec.III in the case of composite  Higgs boson  the  cross section  decreases when $\alpha \rightarrow 1$ (see the arrow depicted in the Fig.4). This fact is a consequence of  a combination of two effects in the cross section:  Decreasing of $\lambda^{T\alpha}_{{}_{3H}}$ due to the increase in $\alpha$ (see Fig.(2)), and the increase of $M^{dyn}_{{}_{H}}$. 

Although enhancement of the signal for Higgs boson pair production is possible for light standard
model scalar bosons\cite{baur}, and even in some extensions of
the standard model\cite{zerwas}, we do not see any possible enhancement of the trilinear 
self-coupling, and, consequently, of the Higgs boson pair production in
the case of heavy dynamically generated scalar bosons. Note that the cross section will
be basically impossible to be measured at the LHC, even if the self-couplings, determined
from Eqs.(\ref{e19}) and (\ref{e20}), are of $O(50)\lambda_{0}$ for the scalar masses that we are considering. There are possible refinements in these calculations, like the introduction of NLO
corrections\cite{spira}, which are not considered because we just want
an order of magnitude estimate, therefore any refinement that may be introduced
will not modify the fact that the production of heavy composite Higgs bosons will
have smaller rates than in the case of fundamental bosons. 

\section{Conclusions} 
  
  In this work we  have presented  a discussion about the general properties of the trilinear self-coupling of a composite Higgs boson based on a general ansatz  for the technifermion self-energy. If the Higgs boson is  composite  we can expect it to be, at least in the most usual models, a very massive particle, $M_{{}_{H}}\propto O(0.6 - 1.2)TeV$, as in the examples of technicolor gauge groups discussed above ($SU(4)_{{}_{TC}}$ or $SU(2)_{{}_{TC}}$). We  verified in the Section IV, assuming in our ansatz the extreme walking technicolor limit,  $\alpha \rightarrow 0$, that the cross section for Higgs boson pair production depicted in the Fig.(4) practically does not differ from the one of the fundamental standard model Higgs boson. The cross section in the limit $\alpha \rightarrow 1$ is very different from the standard result. According to what is known for a long time, this dynamics is exactly the one that leads to the many phenomenological problems in technicolor models. In conclusion, if the origin of the gauge symmetry breaking of the standard model is a dynamical one, {\it i.e.} the scalar Higgs boson is a composite one, it will be much harder than it is in the case of the standard model (with fundamental scalar bosons) to obtain information about the mechanism with the study of the trilinear self-couplings at the LHC. The best possibility to study this coupling, at least for the models usually discussed in the literature, will happens in the case of an extreme walking technicolor dynamics.

\section*{Acknowledgments}

This research was supported by the Conselho Nacional de Desenvolvimento
Cient\'{\i}fico e Tecnol\'ogico (CNPq) (AAN) and by Fundac\~ao de Amparo \`a
Pesquisa do Estado de S\~ao Paulo (FAPESP) (AD).

\begin{figure}[t]
\begin{center}
\epsfig{file=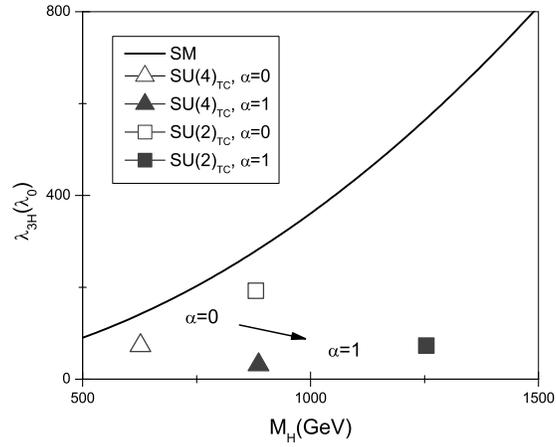,width=0.5\textwidth}
\caption{Trilinear couplings as a function of the Higgs mass for a fundamental and composite Higgs boson. }
\end{center}
\end{figure}
\begin{figure}[bt]
\begin{center}
\epsfig{file=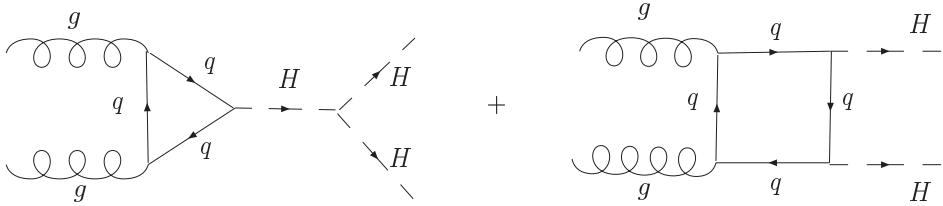,width=0.8\textwidth}
\vspace{0.3cm}
\caption{Diagrams that contribute to the process $gg \rightarrow HH$. For convenience we omitted the crossed terms.}
\end{center}
\end{figure}
\begin{figure}[bt]
\begin{center}
\epsfig{file=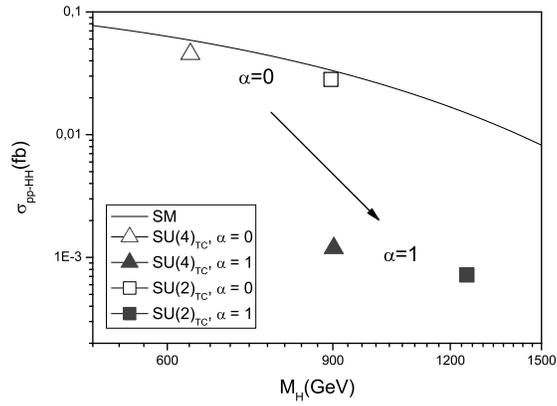,width=0.5\textwidth}
\caption{Total cross section for Higgs boson pair production via gluon-fusion mechanism. The figure was computed
assuming $M_{{}_{H}}\gg m_{q}$. The curve is the result for the standard model fundamental Higgs boson, and the
small box and triangles indicate the value of the cross sections for different technicolor groups with different
dynamics (values of the parameter $\alpha$).}
\end{center}
\end{figure}

\end{document}